\documentclass[lettersize,journal]{IEEEtran}
\usepackage{amsmath,amsfonts}
\usepackage{algorithmic}
\usepackage{array}
\usepackage[caption=false,font=normalsize,labelfont=sf,textfont=sf]{subfig}
\usepackage{textcomp}
\usepackage{stfloats}
\usepackage{url}
\usepackage{verbatim}
\usepackage{graphicx}
\def\BibTeX{{\rm B\kern-.05em{\sc i\kern-.025em b}\kern-.08em
    T\kern-.1667em\lower.7ex\hbox{E}\kern-.125emX}}
\usepackage{cite}
\usepackage{hyperref} 
\usepackage{color}   
\usepackage{diagbox}  
\usepackage{tabularx}

\begin{document}
\title{Deterministic Computing Power Networking: Architecture, Technologies and Prospects} 
\author{
\IEEEauthorblockN{
Qingmin~Jia\IEEEauthorrefmark{1},
Yujiao Hu \IEEEauthorrefmark{1},
Xiaomao Zhou \IEEEauthorrefmark{1},
Qianpiao Ma \IEEEauthorrefmark{1},
Kai Guo \IEEEauthorrefmark{1},
Huayu Zhang \IEEEauthorrefmark{1},
Renchao Xie\IEEEauthorrefmark{1}\IEEEauthorrefmark{2},\\
Tao Huang\IEEEauthorrefmark{1}\IEEEauthorrefmark{2}
and Yunjie Liu\IEEEauthorrefmark{1}\IEEEauthorrefmark{2}
\\}
\IEEEauthorblockA{\IEEEauthorrefmark{1} Purple Mountain Laboratories, Nanjing 211111, China,\\}
\IEEEauthorblockA{\IEEEauthorrefmark{2} Beijing University of Posts and Telecommunications, Beijing 100876, China \\}

}


\maketitle

\begin{abstract}

With the development of new Internet services such as computation-intensive and delay-sensitive tasks, the traditional ``Best Effort" network transmission mode has been greatly challenged. The network system is urgently required to provide end-to-end transmission determinacy and computing determinacy for new applications to ensure the safe and efficient operation of services. Based on the research of the convergence of computing and networking, a new network paradigm named deterministic computing power networking (Det-CPN) is proposed. 
In this article, we firstly introduce the research advance of computing power networking. And then the motivations and scenarios of Det-CPN are analyzed. Following that, we present the system architecture, technological capabilities, workflow as well as key technologies for Det-CPN. Finally, the challenges and future trends of Det-CPN are analyzed and discussed.

\end{abstract}


\section{Introduction}

With the development of emerging network applications such as artificial intelligence (AI), autonomous driving, cloud virtual reality (VR) and intelligent manufacturing, these new applications have put forward higher requirements for network transmission latency and computing power. For example, the GPT-3 model has 175 billion parameters, and training the GPT-3 model requires 355 GPU years (a GPU V100 runs for 355 years) \cite{lu2021learning}. According the research report \cite{huawei2030comp}, it is estimated that by 2030, total general computing power will see a tenfold increase and reach 3.3 ZFLOPS, and AI computing power will increase by a factor of 500, to more than 100 ZFLOPS.
In addition, in the filed of industrial automation, the communication between Programmable Logic Controllers (PLCs) usually has requirements regarding the upper bound of latency, and the underlying networking infrastructure must ensure a maximum end-to-end message delivery time in the range of 100us to 50ms \cite{grossman2019deterministic}.
Hence, it is necessary and significant to design a new network architecture with ultra-low latency, ultra-high bandwidth, and ultra-strong computing power. 

To cope with the challenges brought by new application development, the academia and industry have been actively exploring. For example, in order to meet the challenges of new business demands for computing power, computing power networking (CPN) has been proposed, aiming at connecting distributed computing nodes, achieving rapid access to computing resources and efficient distribution of computing tasks\cite{tang2021computing}. Meanwhile, in order to address the challenge of latency and jitter requirements for new services, deterministic networks has also been proposed, aiming at ensuring the quality and reliability of data transmission\cite{huang2022cycle}. 

However, some emerging network applications (e.g., Cloud VR, autonomous driving) have both latency-sensitive and computation-intensive characteristics, putting forward high demands on both latency and computing power. And the current research on CPN mainly focuses on how to schedule computing tasks to matching computing nodes, while neglecting the transmission determinacy and computation determinacy. Therefore, the current CPN cannot solve the determinacy problems of transmission and computation. How to design a new network architecture that meets latency and computing power requirements has become a major challenge at present.

Fortunately, many researchers have begun to pay attention to this issue.
In \cite{peng2022enabling}, the authors propose a task deterministic network architecture that provides communication with bounded low latency and zero jitter for critical tasks among edge computing systems. 
In \cite{zhang2023detcncs}, the authors proposed a deep reinforcement learning based deterministic scheduling architecture for computing and networking convergence, achieving deterministic end-to-end transmission with bounded latency.
However, these works usually only considered transmission determinacy, and did not consider computing determinacy. 
As a result, it is still difficult to meet the demands of time-sensitive and computation-intensive computing tasks.

In this paper, we propose a new network paradigm called deterministic computing power networking (Det-CPN), which is based on computing power networking and deterministic network technology, and integrates deeply network and computing resources.  
The Det-CPN can provide end-to-end deterministic computing-network service capabilities for time-sensitive and computation-intensive applications, achieving the determinacy for latency, jitter, path, and computing.
Thus, the Det-CPN can effectively address the challenges that traditional ``Best Effort" network transmission mode and ``Time Division and Sharding" node computing method cannot solve, and can meet the needs of new business development.
The main contributions of this article are highlighted as follows:

$\bullet$ The motivations and scenarios for Det-CPN is analyzed. Computation-intensive and time-sensitive application has equally important requirements for deterministic transmission and deterministic computation.

$\bullet$ An architecture of Det-CPN with network determinacy and computing determinacy is proposed for time-sensitive and computation-intensive applications. And the technological capabilities and workflow of Det-CPN is presented. 
 
$\bullet$ The key technologies of Det-CPN are introduced, including network determinacy technology and computing determinacy technology.

$\bullet$ The challenges and future trends of Det-CPN are analyzed and discussed.

The remainder of the article is organized as follows. We provide the overview of CPN, and analyze the motivations and scenarios for Det-CPN. Then, the architecture of Det-CPN is proposed, and the technological capabilities, workflow as well as the key technologies are presented. Following that, some research challenges and future trends are discussed. Finally, we conclude the article.

\section{Research Advance for CPN}

Det-CPN can be considered as the next evolution paradigm of CPN. In order to better understand Det-CPN, we summarize and analyze the research progress of CPN in this section. By improving the design of network architecture and protocols, CPN connects distributed computing nodes and coordinates scheduling, achieving performance optimization as well as efficient utilization of computing-network resources. 
Currently, CPN has achieved initial results in system architecture, technological innovation, and standard specifications. In terms of architecture, the current mainstream architecture design of CPN includes centralized architecture and distributed architecture. 

In the centralized architecture scheme, the CPN is divided into control plane and data plane. The control plane has a full resource view of the CPN, and makes unified computing power scheduling decisions. Through mechanisms such as centralized computing-network scheduling, it achieves routing addressing, distribution scheduling, and resource allocation of computing network resources. 
Based on a centralized approach, a large amount of research has been conducted in both academia and industry.
In \cite{yang2023skypilot}, the authors proposed the concept of Sky Computing, which obtains resource service status information of distributed computing nodes through centralized methods, and then performs global unified task scheduling.
Moreover, the authors in \cite{10288527} proposed a Service Intent-aware Task Scheduling framework for CPN to achieve the optimal matching of task intent and computing-networking resources. 

In the distributed architecture scheme, the CPN achieves synchronization of the status information through the interaction between adjacent routing nodes. And the routing and forwarding of computing tasks are also completed during the decision-making of network nodes. And this scheme is usually implemented through network layer protocol extension. 
Based on a distributed approach, the academic community has also conducted extensive research on CPN. 
In \cite{krol2019compute}, the authors proposed the Compute First Networking technology solution, which is implemented using the Border Gateway Protocol (BGP) extension method. And it implements the synchronization of computing-network status information, routing and forwarding decisions of computing tasks at the network layer.
In \cite{liu2021cfn}, the authors designed a scheduling strategy based on load balancing to allocate users' computing power tasks to an optimal computing power site by sensing the load and network status of each computing site.

The standardization of CPN has also made significant progress. 
The International Telecommunication Union (ITU) in July 2021 approved CPN standards such as Y.2501 "Computing Power Network framework and architecture" \cite{itu2021cpn}, marking new progress in the internationalization of CPN standards. The Internet Engineering Task Force (IETF) has also established the Computing in the Network Research Group (COINRG) working group to carry out research and standardization work on intra network computing.

\section{Motivations and Scenarios for Det-CPN}

Det-CPN has important application value in emerging business fields that are time-sensitive and computation-intensive. In this section, we take intelligent driving, Cloud VR and intelligent manufacturing as examples to analyze the motivations and typical application scenarios of Det-CPN.

\subsection{Intelligent driving}

Intelligent driving is a core application service in future intelligent society. It relies on the vehicle's own cameras, millimeter wave radar, LiDAR, inertial navigation and other sensors for environmental perception, and then performs calculations, decisions, and control execution, which requires strong computing power support and strict low latency communication guarantees. However, massive perceptual data and complex computing tasks make it difficult to process at a lower cost on vehicle computing platforms. Therefore, the intelligent driving will move towards the trend of ``cloud-network-edge-end" integrated development. The cloud, network, edge, and vehicle end are connected and integrated through Det-CPN, achieving a flexible, agile, timely and accurate deterministic computing-network environment and resource supply model. Det-CPN can achieve collaborative scheduling of computing tasks based on the latency requirements of applications. 

\subsection{Cloud VR}

Cloud VR refers to the introduction of the concepts and technologies of cloud computing and cloud rendering into VR business applications. 
Compared to traditional VR, Cloud VR has the advantages of reducing user terminal costs, improving VR resource utilization efficiency, and facilitating centralized content management in the cloud. However, Cloud VR has relatively high requirements for computing and network performance. Usually, the motion-to-photon (MTP) latency cannot exceed 20 ms. At the same time, the typical computing requirements include 8K H.265 real-time hard decoding and multi-channel parallel computing capabilities \cite{chen2023multi}. Users may experience dizziness if their viewing experience is repeatedly hindered by excessive latency. In order to avoid dizziness, the strong interactive services of Cloud VR require deterministic latency and jitter. Therefore, Cloud VR requires the support of Det-CPN.


\subsection{Intelligent Manufacturing}

With the development of the manufacturing industry towards intelligent and digital transformation, industrial control systems are gradually moving towards cloud deployment, and the process operations on the production site can be remotely controlled and processed to ensure production flexibility and safety. At the same time, the cloud deployment of intelligent manufacturing also allows large enterprises to achieve production factor allocation and optimization between headquarters and multiple bases on a larger scale, achieving cost reduction and efficiency increase for enterprises. Therefore, in response to the trend of industrial control systems towards wide area and cloud development, Det-CPN can provide real-time computing power and real-time transmission guarantee for the next generation of industrial control systems. For example, deploying the factory control system in the form of cloud services on the cloud, transmitting the information collected by sensing devices with ultra-low latency and ultra-high reliability to edge computing nodes. Through rapid identification and decision-making, control instructions are quickly fed back to terminal devices and actions are executed.

\section{The architecture, technological capabilities and workflow of Det-CPN} 

\begin{figure*}[htb]
\centering
\includegraphics[width=\linewidth]{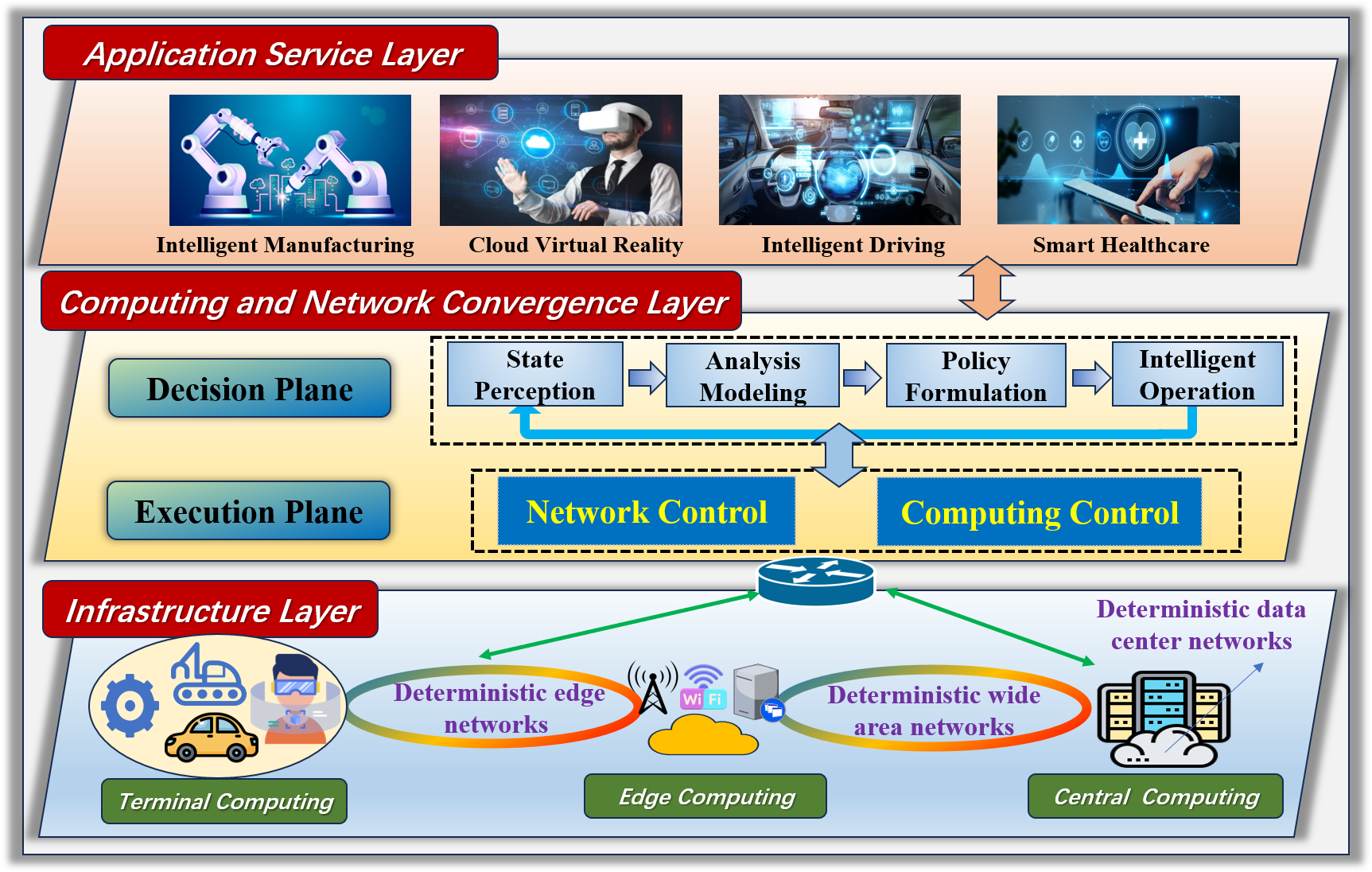}
\caption{The architecture of Det-CPN}
\label{fig_arch}
\end{figure*}

Det-CPN is an advanced stage in the development of CPN, fully considering the time constraints of computing tasks in terms of transmission and computation. By adopting deterministic mechanism methods, deterministic transmission and computation of computing tasks within the constraint time are achieved. In this section, we present the architecture, analyze the technological capabilities and introduce the workflow for Det-CPN system.

\subsection{Architecture Design of Det-CPN}

In this section, we present the architecture of Det-CPN. As shown in Fig.\ref{fig_arch}, the architecture mainly consists of three parts, namely the infrastructure layer, computing and network convergence layer, and application service layer.

\subsubsection{Infrastructure Layer}

The infrastructure layer includes heterogeneous multi-level computing infrastructure and heterogeneous ubiquitous network infrastructure. 
Computing infrastructure provides heterogeneous computing power resources such as basic computing power, intelligent computing power and super computing power, usually including terminal computing nodes, edge computing nodes and central cloud computing nodes. 
Network infrastructure provides end-to-end network connectivity, including deterministic edge networks, deterministic wide area networks, and deterministic data center networks. The infrastructure layer provides basic computing and network resource capabilities for the upper layer.

\subsubsection{Computing and Network Convergence Layer}

The computing and network convergence layer is the core of the Det-CPN, including the computing and network decision plane (CNDP), the computing and network execution plane (CNEP).

The CNDP consists of state perception, analysis modeling, policy formulation, and intelligent operation.
The CNDP obtains computing-network information through the functions of the state perception module, such as state detection, task perception, network perception, and intention perception. And then based on the state information, the analysis modeling module can achieve computing measurement, computing modeling, task deconstruction, and knowledge modeling. On the basis of analysis and modeling, policies such as network configuration, application orchestration, swarm intelligence learning, and intelligent distribution are formulated through the policy formulation module. At the same time, the CNDP achieves integrated services, intention driven decision-making, adaptive optimization, and autonomous operation and maintenance through the intelligent operation module. The CNDP will hand over the formulated strategies to the CNEP for execution.

The CNEP is divided into network control module and computing control module. 
The network control module mainly controls the network infrastructure, including edge networks, wide area networks, and data center networks, so as to achieve end-to-end deterministic and high-quality network control. 
The computing control module mainly manages and orchestrates the multi-level computing resources, achieving functions such as hierarchical and domain based computing, heterogeneous computing, computation offloading, and serverless scheduling. Its core capability is the deterministic computing processing of computing tasks. Namely, it constrains the processing delay of computing tasks to a fixed range, thereby ensuring that user terminal can obtain the results of computing task completion and return within the constraint time.

\subsubsection{Application Service Layer}

The application service layer provides users with various application services and is responsible for service operation. 
Application services mainly include computation-intensive and time-sensitive applications such as Cloud VR, intelligent driving, intelligent manufacturing, and smart healthcare.

\subsection{The Technological Capabilities of Det-CPN}

In order to achieve the design goals of Det-CPN, it is necessary to possess three core capabilities, namely, deterministic transmission capability, deterministic computing capability, intelligent decision-making capability. In this subsection, we present these three technical capabilities.

\subsubsection{ Deterministic transmission capability}

In Det-CPN, the transmission of computing tasks usually has strict limitations on latency and jitter. Therefore, providing end-to-end deterministic transmission guarantee for computing tasks is one of the core capabilities of Det-CPN. On the edge network side, the deterministic edge network is constructed by introducing time sensitive networking (TSN) and ``5G+TSN" technology. On the wide area network side, the deterministic wide area network is implemented by introducing deterministic networking (DetNet) and segment routing (SR) technology. On the network side of the data center, the deterministic data center network introduces technologies such as intelligent lossless networks. And combined with software defined network (SDN) technology, the Det-CPN can achieve end-to-end deterministic transmission, ensuring latency determinacy, jitter determinacy, and path determinacy.

\subsubsection{ Deterministic computing capability}

In computation-intensive and delay-sensitive applications, the total transmission and processing time of computing tasks are constrained, only ensuring transmission determinacy cannot meet the delay requirements of some applications. Therefore, in order to ensure ultra-low latency in end-to-end transmission and computation, and prevent incoming computing tasks from queuing up, it is necessary to promptly process computing tasks that arrive at computing processing units (such as CPUs and GPUs). If the traditional ``time division and sharding" computing method is used, it will be difficult to ensure the processing delay of the computing tasks. Therefore, by designing mechanisms such as prioritization of computing tasks, preemption of high priority tasks, reservation and locking of computing resources, and elastic scaling of computing resources, deterministic computing can be achieved. It should be emphasized that computational determinacy refers to the time required to complete computing tasks within a bounded time range.

\subsubsection{Intelligent decision-making capability}

Intelligentization is the future development trend of CPN. Det-CPN also needs to possess the ability of network intelligence, and based on artificial intelligence strategies and approaches, to achieve self perception, self configuration, self optimization, self decision-making, and self maintenance. Specifically, the CNDP of Det-CPN needs to be based on intelligent strategy methods to achieve functions such as optimizing network service perception, computing-network task scheduling, and computing-network resource orchestration. On the other hand, considering that single point intelligence cannot meet the development of future computing-network intelligence services, Det-CPN also needs to have intelligent networking capabilities, that is, by interconnecting and sharing intelligent models, intelligent resources, etc., to improve the overall intelligence level of the computing power networks.

\subsection{The Workflow of Det-CPN}

\begin{figure}
\centering
\includegraphics[width=\linewidth]{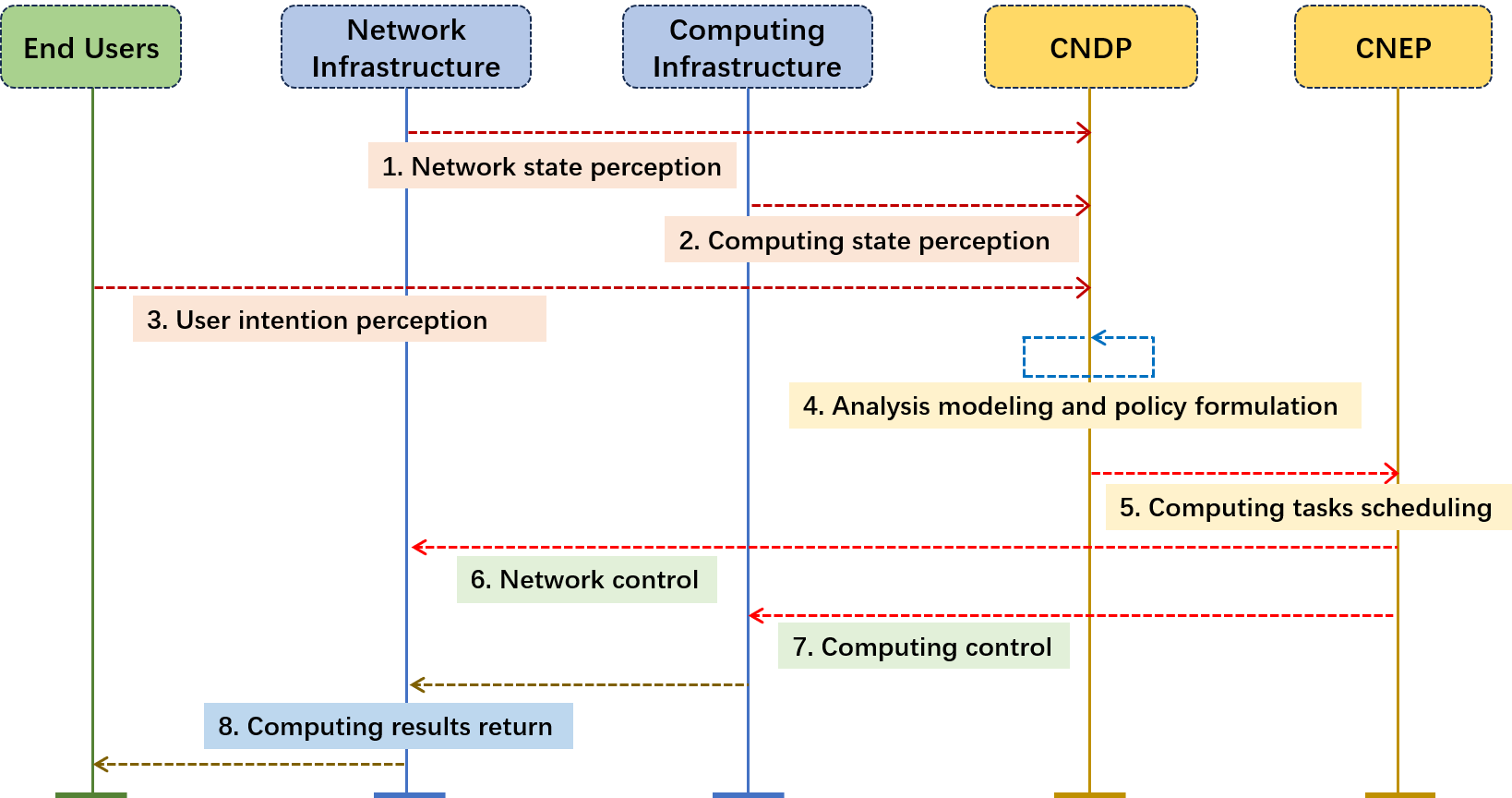}
\caption{The workflow of Det-CPN}
\label{fig_workfolw}
\end{figure}

In order to further understand the working mechanism of Det-CPN, we present the basic workflow of Det-CPN in this subsection, as shown in Fig.\ref{fig_workfolw}.

\textbf{Step1:} Network state perception. The CNDP system collects the status information of network resources, then stores them in the status information database and updates them regularly.

\textbf{Step2:} Computing state perception. The CNDP system collects the status information of computing resources and computing services, then stores them in the status information database and updates them regularly.

\textbf{Step3:} User intention perception. The CNDP system also needs to collect user intention information, including latency, jitter, packet loss rate, computing power type, etc., so as to better provide customized computing-network services.

\textbf{Step4:} Analysis modeling and policy formulation. Based on computing-network state information and user intention information, the CNDP system will conduct analysis, modeling, and policy formulation.

\textbf{Step5:} Computation tasks scheduling. The CNDP system obtains the optimal computing node and transmission path through perception and analysis.

\textbf{Step6:} Network control. According to the decision results of CNDP, the network controller in CNEP needs to route and distribute computation tasks to the target computing nodes. 

\textbf{Step7:} Computing control. The computing controller in CNEP adopts deterministic computing technology to efficiently handle computation tasks.

\textbf{Step8:} Computing results return. After the computing node completes the task processing, the computing results are returned to the user.

\section{Key Technologies of Det-CPN}

Det-CPN needs to integrate multiple key network technologies to achieve its goals, including edge network determinacy, wide area network determinacy, data center network determinacy, as well as computing determinacy on computing nodes. Therefore, in this section, we analyze and discuss the key technologies of the Det-CPN in detail.


\subsection{Network Determinacy Technology}

Network determinacy technology adopts technical mechanisms such as delay determinacy, jitter determinacy, and path determinacy, to construct an end-to-end deterministic network system, which is a key enabling technology for Det-CPN. In this paper, we divide the network determinacy technology into edge network determinacy, wide area network determinacy, and data center network determinacy. 

\subsubsection{Edge Network Determinacy}
\

Edge network determinacy mainly refers to the access side network determinacy, and is the foundation for achieving deterministic computing power network access. It usually includes wireless edge networks and wired edge networks.

The determinacy of wireless edge networks usually requires 5G access network technology, including gigabit bandwidth access capability and millisecond level highly reliable transmission capability. In addition, Det-WiFi can also be applied to wireless edge networks, which support high-speed applications and provide better deterministic services in practical multi-hop edge environments.
The determinacy of wired edge networks is mainly based on TSN technology. TSN technology mainly focuses on the network link layer, with mechanisms such as clock synchronization, traffic shaping, resource reservation, path selection, and fault tolerance to ensure deterministic latency.
Moreover, there is a demand for wireless and wired hybrid deterministic networking between edge computing nodes, and ``5G+TSN" is considered as an effective candidate for deterministic edge network. Combined with the precise clock synchronization ability and deterministic traffic scheduling ability of TSN technology, edge network can ensure low latency and highly reliable transmission of various business flows, thereby providing high-quality and highly reliable edge network support. 


\subsubsection{Wide Area Network Determinacy}
\

Wide area network determinacy technology is a new network technology aimed at large-scale and long-distance transmission, which provides deterministic service quality such as low latency, low jitter, low packet loss rate, high bandwidth, and high reliability. 
This technology is a collection of a series of protocols and mechanisms, achieving deterministic latency through mechanisms such as clock synchronization, frequency synchronization, scheduling shaping, and resource reservation. The deterministic jitter and packet loss rate can be implemented through mechanisms such as priority division, jitter reduction, and buffer absorption. And the deterministic reliability is achieved through technologies such as multiplexing, packet replication and elimination, and redundant backup. The representative technology for wide area network determinacy includes deterministic networking (DetNet) and Segment Routing (SR).

DetNet focuses on the network layer and defines deterministic algorithms for traffic queuing, shaping, scheduling, and preemption. By defining traffic control rules, it achieves deterministic jitter and packet loss rate. 
In Det-CPN, the use of deterministic networking technology enables the determination of latency and jitter in computing task transmission.
The deterministic and non deterministic services can be flexibly switched, and the level of deterministic service quality can be autonomously controlled. It can provide deterministic data transmission channels for computing task distribution and enable computing power collaboration across nodes, clusters, and regions. SR is a source packet routing technique in which network nodes forward data packets based on an ordered list of instructions called segments. The segmented routing mechanism simplifies traffic engineering and management across network domains. 
Applying SR technology in Det-CPN can monitor the entire network topology and its traffic in real-time, and based on these data, determine the network transmission paths that computing tasks should pass through, and allocate bandwidth to these paths. Therefore, in Det-CPN, SR technology is used to achieve path determinacy, ensuring accurate distribution and transmission of computing tasks.

\subsubsection{Data Center Network Determinacy}
\

The data center network (DCN) is usually the ``Last Mile" of the end-to-end deterministic networks, which plays an important role in Det-CPN. DCN determinacy refers to that the DCN has deterministic low latency capabilities, and it usually requires lossless network technology to provide a low latency and high throughput network environment for Det-CPN, thereby accelerating the efficiency of computing and storage, and greatly improving user experience. 
Lossless network is a new type of low latency network that deeply integrates computing, storage, and network to improve and innovate in congestion control, flow control, packet forwarding, routing, and other aspects of the network. 
Lossless network can provide a ``zero packet loss, low latency and high throughput" network environment for Det-CPN, combined with intelligent lossless algorithms, thereby improving the performance of DCN. In addition, lossless network can be combined with other technologies to further expand their role in Det-CPN. For example, in combination with Remote Direct Memory Access (RDMA) technology, applications are used to directly read or write to remote memory, avoiding the intervention of operating systems and protocol stacks, achieving more direct, simple, and efficient data transmission, significantly reducing the time required in the data transmission process. 

\subsection{Computing Determinacy Technology}

With the deployment of time-sensitive and computation-intensive applications in the cloud, CPN has increasingly high requirements for the computing determinacy. Det-CPN needs to ensure that computing tasks return execution results within the required latency of user terminals, subject to time constraints of end user business. Therefore, the traditional ``time division and sharding" cloud computing method will no longer be applicable to Det-CPN scenarios. Deterministic computing technology needs to provide deterministic service quality with low computing latency, low latency jitter, high reliability, etc. It needs to provide stable and controllable deterministic service quality assurance for different businesses, and have the ability to respond to sudden large amounts of computing requests. To achieve deterministic computing, it is necessary to deeply improve the current CPN based on the key technologies such as task grading, resource reservation, and resource pre-adjustment.

\subsubsection{Task Grading Mechanism}
\

In computing determinacy, computing task requests can be divided into multiple priorities based on application latency requirements and computing load characteristics. Computing task requests with the requirements of lower latency and higher computational load should have higher processing priority. When task requests with different priorities arrive, high priority requests should be given priority in allocating computing resources to meet their latency requirements. When allocating resources for higher priority requests, it should be allowed to adjust the resource allocation strategy for lower priority requests and allow higher priority task requests to preempt computing resources.

\subsubsection{Resource Reservation Mechanism}
\

In computing determinacy, the computing resources of nodes should be reserved to respond to sudden computing requests. 
When computation-intensive and delay-sensitive task requests arrive at the computing node, task requests can be immediately allocated computing resources without the need to queue and wait. Due to the existence of redundant computing resources in computing power nodes, the performance of real-time processing for computing tasks will be greatly improved.

\subsubsection{Resource Pre-adjustment Mechanism}
\

In computing determinacy, serverless computing technology can be applied for server scaling to respond to sudden computing requests. However, due to the current serverless computing technology, which is mostly based on server load level for dynamic scaling, the additional delay caused by cold start during scaling can affect real-time computing. The lagging scaling strategy during scaling will lead to unnecessary waste of network computing resources. Therefore, pre-adjustment of computing resources can be based on network resource situational awareness and task request prediction technology. Namely, when an increase in task requests is predicted, computing resources can be pre-deployed to the hot pool through cold start. When task requests increase, hot start can be used to avoid startup delay. When task requests are reduced, a pre-shrinking mechanism can be established to optimize computing resources. When the number of task requests drops to a certain threshold, timely recycling of partial computing resource allocation.

\section{The Challenges and Future Trends of Det-CPN}

Det-CPN paves the way for end-to-end deterministic transmission and deterministic computing, while leaving some challenges
to be discussed. In this section, we analyze some potential research issues that need to be discussed for future research.

\subsection{Integrated Modeling and Control for Computing and Networking}

Det-CPN is a new paradigm that combines network transmission and computing processing, requiring unified scheduling and control for the networks and computing. However, due to the fact that networks typically belong to communication network operators and computing power typically belongs to cloud service providers, the management and control of networks and computing power are separated, making it difficult to achieve integrated scheduling and control of computing-network resources. Therefore, it is necessary to build a unified Det-CPN operating system on top of communication network operators and cloud service providers, jointly model and control the network and computing resources, plan the total latency of task transmission and task calculation, so as to meet the transmission and computing latency requirements of service businesses.

\subsection{Integration of forwarding, computing and caching}

Currently, Det-CPN solution adopts an Overlay approach, which controls the computing and network resources of the infrastructure layer to achieve deterministic transmission and computing processing of computing tasks, without making fundamental modifications to existing network devices. Facing the future, in order to reduce the transmission and processing latency of computing tasks, processing computing tasks in network devices may become a trend. The devices in Det-CPN will have forwarding, computing, and caching capabilities, namely, the ability of integration of forwarding, computing and caching. When the network device receives a computing task, it will use the computing resources integrated by the network device for task calculation processing and cache the results. The cached results can be used to simplify the computational cost of similar computing tasks. By deploying network devices that integrate computing and storage in network edge environments, the processing and response latency of computing tasks can be greatly reduced, meeting the new business requirements in Det-CPN.

\subsection{Transmission Control Optimization}

In Det-CPN, both network and computation adopt deterministic related technologies, greatly improving the reliability of data transmission and processing for computing tasks. Due to the current network adopting a layered design approach, the transport layer is decoupled from the network layer and link layer, and the transport layer is insensitive and untrustworthy of the underlying network conditions. Therefore, Transmission Control Protocol (TCP) is commonly used in the transmission layer to ensure the reliability of data transmission by increasing the complexity of the network protocol stack. With the adoption of technologies such as deterministic networks in Det-CPN, data transmission has highly reliable, high-quality, and predictable capabilities, and the traditional transmission control protocol mechanisms becomes redundant. Therefore, it is necessary to design new transmission control optimization methods, cut out complex and redundant traditional transmission control mechanisms or redesign them to ensure efficiency, simplicity and lightweighting. The transmission control optimization of Det-CPN can be achieved through technologies such as fine-grained congestion detection, network traffic prediction.

\section{Conclusion}

Providing end-to-end transmission and computing determinacy for computation-intensive and time-sensitive applications is of great significance. In this article, on the basis of research on CPN, we presented the architecture, technological capabilities and workflow of Det-CPN, and analyzed its application scenarios, key technologies. Finally, the potential technical research challenges and future trends were discussed.

\section{Acknowledgment}

This work was supported by the Jiangning Baijia Lake Plan Program (No.74072203-3),  the National Natural Science Foundation of China (No. 92267301).
\bibliographystyle{IEEEtran}
\bibliography{reference}

\begin{thebibliography}{10}
\providecommand{\url}[1]{#1}
\csname url@samestyle\endcsname
\providecommand{\newblock}{\relax}
\providecommand{\bibinfo}[2]{#2}
\providecommand{\BIBentrySTDinterwordspacing}{\spaceskip=0pt\relax}
\providecommand{\BIBentryALTinterwordstretchfactor}{4}
\providecommand{\BIBentryALTinterwordspacing}{\spaceskip=\fontdimen2\font plus
\BIBentryALTinterwordstretchfactor\fontdimen3\font minus \fontdimen4\font\relax}
\providecommand{\BIBforeignlanguage}[2]{{%
\expandafter\ifx\csname l@#1\endcsname\relax
\typeout{** WARNING: IEEEtran.bst: No hyphenation pattern has been}%
\typeout{** loaded for the language `#1'. Using the pattern for}%
\typeout{** the default language instead.}%
\else
\language=\csname l@#1\endcsname
\fi
#2}}
\providecommand{\BIBdecl}{\relax}
\BIBdecl

\bibitem{lu2021learning}
L.~Lu, P.~Jin, G.~Pang, Z.~Zhang, and G.~E. Karniadakis, ``Learning nonlinear operators via deeponet based on the universal approximation theorem of operators,'' \emph{Nature machine intelligence}, vol.~3, no.~3, pp. 218--229, 2021.

\bibitem{huawei2030comp}
{Huawei Technology Report}, ``Computing 2030,'' Tech. Rep., 4 2023.

\bibitem{grossman2019deterministic}
E.~Grossman, ``Deterministic networking use cases,'' \emph{IETF RFC 8578}, 2019.

\bibitem{tang2021computing}
X.~Tang, C.~Cao, Y.~Wang, S.~Zhang, Y.~Liu, M.~Li, and T.~He, ``Computing power network: The architecture of convergence of computing and networking towards 6g requirement,'' \emph{China communications}, vol.~18, no.~2, pp. 175--185, 2021.

\bibitem{huang2022cycle}
Y.~Huang, S.~Wang, T.~Huang, and Y.~Liu, ``Cycle-based time-sensitive and deterministic networks: Architecture, challenges, and open issues,'' \emph{IEEE Communications Magazine}, vol.~60, no.~6, pp. 81--87, 2022.

\bibitem{peng2022enabling}
G.~Peng, S.~Wang, Y.~Huang, T.~Huang, and Y.~Liu, ``Enabling deterministic tasks with multi-access edge computing in 5g networks,'' \emph{IEEE Communications Magazine}, vol.~60, no.~8, pp. 36--42, 2022.

\bibitem{zhang2023detcncs}
W.~Zhang, R.~Guo, D.~Yang, and C.~Zhang, ``Detcncs: Deterministic computing and networking convergence scheduling,'' in \emph{Proc. the ACM Turing Award Celebration Conference-China 2023}, 2023, pp. 59--60.

\bibitem{yang2023skypilot}
Z.~Yang, Z.~Wu, M.~Luo, W.-L. Chiang, R.~Bhardwaj, W.~Kwon, S.~Zhuang, F.~S. Luan, G.~Mittal, S.~Shenker \emph{et~al.}, ``Skypilot: An intercloud broker for sky computing,'' in \emph{Proc. USENIX NSDI}, 2023, pp. 437--455.

\bibitem{10288527}
Q.~Tang, R.~Xie, L.~Feng, F.~R. Yu, T.~Chen, R.~Zhang, and T.~Huang, ``{SIaTS}: A service intent-aware task scheduling framework for computing power networks,'' \emph{IEEE Network}, pp. 1--1, 2023.

\bibitem{krol2019compute}
M.~Kr{\'o}l, S.~Mastorakis, D.~Oran, and D.~Kutscher, ``Compute first networking: Distributed computing meets icn,'' in \emph{Proc. ACM ICN}, 2019, pp. 67--77.

\bibitem{liu2021cfn}
B.~Liu, J.~Mao, L.~Xu, R.~Hu, and X.~Chen, ``{CFN-dyncast}: Load balancing the edges via the network,'' in \emph{Proc. IEEE WCNC Workshops}, 2021, pp. 1--6.

\bibitem{itu2021cpn}
{ITU}, ``Computing power network- framework and architecture: Y.2501.''\hskip 1em plus 0.5em minus 0.4em\relax ITU, 2021.

\bibitem{chen2023multi}
L.~Chen, Y.~Tang, J.~Xia, S.~Chen, C.~Zheng, H.~Lin, and W.~Wang, ``{Multi-MEC collaboration for VR video transmission: Architecture and cache algorithm design},'' \emph{Computer Networks}, vol. 234, p. 109864, 2023.

\end{thebibliography}
\begin{IEEEbiographynophoto}{Qingmin Jia}
received the Ph.D. degree from Beijing University of Posts and Telecommunications in 2019. He is currently a Researcher at Purple Mountain Laboratories. His current research interests include Computing and Network Convergence, and deterministic networks.
\end{IEEEbiographynophoto}\vspace{-38pt}
\begin{IEEEbiographynophoto}{Yujiao Hu}
currently works in the Purple Mountain Laboratories. She obtained her Bachelor and PhD degrees from the Department of Computer Science of Northwestern Polytechnical University of China in 2016 and 2021 respectively. She focuses on deep learning, edge computing, multi-agent cooperation. 
\end{IEEEbiographynophoto}\vspace{-38pt} 
\begin{IEEEbiographynophoto}{Xiaomao Zhou}
received his Ph.D. degree from Harbin Engineering University in 2020. He is currently a faculty member at Purple Mountain Laboratories. His current research interests include edge intelligence, and autonomous computing and network convergence.
\end{IEEEbiographynophoto} \vspace{-38pt} 
\begin{IEEEbiographynophoto}{Qianpiao Ma}
received his Ph.D. degree from the University of Science and Technology of China in 2022. He is currently a post-doctoral researcher at Purple Mountain Laboratories. His research interests include edge computing and distributed machine learning.
\end{IEEEbiographynophoto} \vspace{-38pt} 
\begin{IEEEbiographynophoto}{Kai Guo}
received his Ph.D. degree from the University of Tsukuba, Japan, in 2020. He is currently a researcher at Purple Mountain Laboratories, China. His research interests include computing and network convergence, mobile computing, and wireless networks.
\end{IEEEbiographynophoto} \vspace{-38pt} 
\begin{IEEEbiographynophoto}{Huayu Zhang}
received the Ph.D. degree from Peking University, in 2017. He is currently a Researcher at Purple Mountain Labs. His research interests include distributed systems, and next generation networks.
\end{IEEEbiographynophoto}\vspace{-38pt} 
\begin{IEEEbiographynophoto}{Renchao Xie}
received the Ph.D. degree from the Beijing University of Posts and Telecommunications, in 2012. He is currently a Professor at Beijing University of Posts and Telecommunications. His research interests include edge computing, and Industrial Internet of Things.
\end{IEEEbiographynophoto}\vspace{-38pt} 
\begin{IEEEbiographynophoto}{Tao Huang}
received his Ph.D. degree from Beijing University of Posts and Telecommunications, in 2007. He is currently a professor at Beijing University of Posts and Telecommunications. His current research interests include network architecture, and deterministic networks. 
\end{IEEEbiographynophoto}\vspace{-38pt} 
\begin{IEEEbiographynophoto}{Yunjie Liu}
received the B.S. degree in technical physics from Peking University, in 1968. He is currently the Academician of the China Academy of Engineering. His research interests include next generation networks, and deterministic networks.
\end{IEEEbiographynophoto}

\end{document}